\documentclass[11pt,a4paper]{article}
\usepackage[utf8]{inputenc}
\usepackage{color}
\usepackage{doi}
\usepackage{amsmath,amssymb,amsfonts}
\usepackage{geometry}
\title{\bf Ermakov-Painlev\'e II Symmetry Reduction of  a Class of Moving Boundary Problems for a \\ Reciprocal Extended mKdV Equation}
\author{
    Colin Rogers$^1$ and Sandra Carillo$^{2, 3}$ \footnote{\bf corresponding author,   email: sandra.carillo@uniroma1.it} \\
    \small $^1$University of New South Wales, Sydney, New South Wales, Australia \\
    \small $^2$ Dipartimento Scienze di Base e Applicate per l'Ingegneria \\
    \small Universit\`a di Roma  ``La Sapienza'', 16, Via A. Scarpa, 00161 Rome, Italy \\
    \small $^3$ Gr. Roma1, IV - Mathematical Methods in NonLinear Physics  (MMNLP)\\
    \small National Institute for Nuclear Physics (I.N.F.N.), Rome, Italy
}
\date{\today}

\begin{document}

\maketitle

\begin{abstract}
A reciprocal transformation is applied to link an integrable extension of the classical solitonic 
mKdV to a novel nonlinear  evolution equation  incorporating a source term. Application of   Ermakov-Painlev\'e II symmetry reduction is made to determine the exact solution to a class of associated moving boundary problems of Stefan-type.
\end{abstract}

\section{Introduction}
Reciprocal transformations as  introduced  in a modern solitonic context in \cite{1} constitute  a class of B\"acklund-type transformations \cite{2,3}  which act on admitted conservation laws.
 In \cite{1}, conjugation was made with the classical nonlinear  superposition principle (permutability theorem) of  Bianchi  as obtained in the theory of  pseudo-spherical surfaces. 
 This allows the iterative algorithmic generation of multi-soliton solutions \cite{3}.
  The linkage via reciprocal transformations of the canonical AKNS and WKI inverse scattering schemes of \cite{4,5} respectively    has been detailed in \cite{6}. Invariance of the 
  $1+1$-dimensional  Dym solitonic hierarchy under a class of reciprocal transformations 
   was established in \cite{7}. In \cite{8}, novel hierarchies of nonlinear evolution equations  were derived which are linked, in turn,  to the canonical  Caudrey-Dodd-Gibbon and, Kaup-Kupershmidt hierarchies. In {  \cite{8c}  as well as in  }\cite{8b}, B\"acklund transformations are applied to study Ermakov and Painlev\'e XXV–Ermakov equations.
Reciprocal Transformations in $2+1$-dimensions as introduced in \cite{9} were shown in \cite{10} to connect the Kadomtsev-Petviashvili, modified  Kadomtsev-Petviashvili and $2+1$-dimensional Dym triad of solitonic   hierarchies.

The application of Painlev\'e II symmetry reduction to derive exact solutions to nonlinear moving boundary problems for a range of canonical solitonic equations was initiated in \cite{11} with an analysis of moving boundary problems of Stefan-type for the solitonic Dym equation \cite{12} and reciprocal associates. This was motivated by a connection to the analysis of the involution of the interface in a Hele-Shaw cell \cite{13}.
In \cite{14}, a solitonic extension of the Dym equation was derived in the geometric context of torsion evolution associated with the { spatial binormal motion} of inextensible curves. This extended Dym equation originated in the analysis of novel peakon solitonic phenomena in hydrodynamics (Camassa and Holm \cite{15}).
In subsequent developments, moving boundary problems of Stefan-type for the solitonic KdV, mKdV and Gardner equations together with reciprocal associates \cite{16, 17, 18} have been shown to be amenable to exact solution via Painlev\'e II  symmetry reduction together with action of reciprocal transformations. In addition, in \cite{19} moving boundary problems of Stefan-type have been solved for a canonical member of the WKI inverse scattering scheme via conjugation of a reciprocal and Möbius type transformation.

Hybrid Ermakov-Painlev\'e II systems were originally derived in \cite{20} in the context of wave packet representations incorporating de Broglie - Bohm potential terms. The canonical base Ermakov-Painlev\'e II equation was derived therein in the analysis of transverse wave propagation in a generalised Moorey-Rivlin hyperelastic material. It has subsequently proved important in a range of physical applications, notably in cold plasma physics \cite{21}, Korteweg capillarity theory \cite{22} and in the analysis of Dirichlet-type boundary value problems for the classical Nernst-Planck electrolysis system \cite{23,24}.
Integrable structure underlying Ermakov-Painlev\'e  II systems has been detailed in \cite{25}. Discretisation of the Ermakov-Painlev\'e  II canonical equation and of associated Dirichlet  and Robin-type boundary value problems has been recently analysed in \cite{26}.

Ermakov-Painlev\'e II   symmetry reduction has been applied in \cite{27} to solve a class of moving boundary problems for an extension of the mKdV equation with novel temporal modulations. Here, a reciprocal transformation is applied to derive the exact solution to an associated nonlinear evolution equation which incorporates a source term.

\section{On a novel reciprocal mKdV equation incorporating a nonlinear source term: Ermakov-Painlev\'e II Symmetry Reductions}

Here, a novel third order nonlinear evolution equation reciprocally related to an S-integrable extension of the canonical solitonic  mKdV equation  is introduced, namely 
\begin{equation}
\Phi_t = \frac{1}{2}(\Phi^{-2})_{xxx} + \Delta_x
\tag{2.1}
\end{equation}
wherein
\begin{equation}
\Delta = \nu  \frac{x^2}{2} + \lambda(t+a)^\mu x^{-4}~~, ~~ \lambda, \mu, \nu \in {\mathbb R}
\tag{2.2}
\end{equation}
Thus, application of the reciprocal transformation
\begin{equation}
dy = \Phi dx + \left[ \frac{1}{2}(\Phi^{-2})_{xx} + \nu  \frac{x^2}{2}   + \lambda(t+a)^\mu x^{-4} \right] dt
\tag{2.3}
\end{equation}
with $\rho = \Phi^{-1}$ yields
\begin{equation}
dx = \rho dy - \rho \left[ \frac{1}{2}(\rho^2)_{xx} + \nu  \frac{x^2}{2}   + \lambda(t+a)^\mu x^{-4} \right] dt
\tag{2.4}
\end{equation}
whence,
\begin{equation}
x_y = \rho, \quad x_t = -\left[\rho_{yy} + \frac{\nu}{2}x^2 \rho + \lambda(t+a)^\mu x^{-4}\rho \right]
\tag{2.5}
\end{equation}
An extension of the classical mKdV equation accordingly results, namely 
\begin{equation}
x_t + x_{yyy} +\frac{\nu}{2} x^2 x_y + \lambda(t+a)^\mu x^{-4} x_y = 0\tag{2.6}
\end{equation}
where, in the sequel,  the parameter  specialisation $\nu = -12$ is adopted.

 Ermakov-Painlev\'e II  symmetry reduction of an extended solitonic mKdV equation of the type (2.6) has recently been detailed in \cite{27} wherein $\mu =-2$ in the temporal modulation term and a symmetry reduction ansatz

\begin{equation}
x = (t+a)^m \Psi\left(\frac{y}{(t+a)^n}\right)
\tag{2.7}
\end{equation}
is applied. On insertion of the latter into (2.6) there result
\begin{equation}
(t+a)^{-3n+1} \Psi''' - 6(t+a)^{2m-n+1} \Psi^2 \Psi' + m \Psi - n \xi \Psi' = 0
\tag{2.8}\end{equation}
with  $\Psi= \Psi(\xi), \xi= y/(t+a)^n$. With $m = -1/3, n = 1/3$, integration of (2.8) yelds
\begin{equation}
\Psi'' - 2\Psi^3 - \frac{1}{3}\xi \Psi - \frac{\lambda}{3}\Psi^{-3} = \zeta ~~,~~ \zeta\in \mathbb{R}.
\tag{2.9}\end{equation}
The  latter, on   appropriate scalings $\Psi=\delta w^*, \xi=\epsilon z$ and with $\zeta=0$  produces the canonical wave Ermakov-Painlev\'e II equation \cite{20}, namely 
\begin{equation}
w^*_{zz} = 2 {w^*}^3 + z w^* + \delta {w^*}^{-3}, \quad \delta \in \mathbb{R}.
\tag{2.10}\end{equation}

The reciprocal link between the extended mKdV equation (2.6) and (2.1) implies that the latter inherits admittance of Ermakov-Painlev\'e II symmetry reduction. In addition, 
it can be embedded in a wide class with the temporal modulation as generated by the application of a class of involutory transformations with  genesis in Ermakov system theory \cite{28}. Ermakov-type temporal modulation of the extended mKdV equation (2.6) is detailed in \cite{27}.

\section{A Class of Reciprocal Moving Boundary Problems}
The preceding Ermakov-Painlev\'e II  symmetry reduction (2.7) was applied in \cite{27} to derive exact solution to a class of moving boundary problems of Stefan-type for the extended mKdV equation (2.6). This class is governed by the nonlinear system

\begin{equation*}
x_t + x_{yyy} - 6x^2x_y + \lambda(t+a)^{-2}x^{-4}x_y =0,
\qquad 0<y<S(t), \qquad t>0,
\end{equation*}
\begin{equation}
\left. \begin{array}{l}
\left.x_{yy}-2x^3-\frac{\lambda}{3}(t+a)^{-2}x^{-3}
=L_m\dot S\,  S^j \right. \\
\qquad\qquad\qquad x=P_mS^{\,j}
\end{array} \right\}\qquad \text{on } y=S(t),\ t>0,\\
\tag{3.1}\end{equation}
\begin{equation*}
\begin{aligned}
\left[
x_{yy}-2x^3-\frac{\lambda}{3}(t+a)^{-2}x^{-3}
\right]_{y=0}
&=H_0(t+a)^k,\qquad t>0,\\
S(0)&=S_0.
\end{aligned}
\end{equation*}
Herein, the moving boundary adopted is
${y=S(t)} = \gamma (t+a)^{1/3} $  whence the initial condition requires that $ S_0=\gamma a^{1/3}$.

The action of the  reciprocal transformation (2.4) with parameters $\mu=-2, \nu=-12$ on the class of extended mKdV moving boundary problems (3.1) yields
\begin{equation*}
\begin{aligned}
\Phi_t=\frac12\left(\Phi^{-2}\right)_{xxx}
+\left[-6x^2+\lambda(t+a)^{-2}x^{-4}\right]_x, \\
x\Big|_{y=0}<x<x\Big|_{y=S(t)}, \qquad t>0
\end{aligned}
\end{equation*}
\begin{equation}
\left. \begin{array}{l}\Phi^{-1}(\Phi^{-1})_x
-2x^3-\frac{\lambda}{3}(t+a)^{-2}x^{-3}
=
L_m S^j,\dot S, \\
\qquad\qquad\qquad x=P_mS^j \end{array} \right\}
\qquad
\text{on }
x\Big|_{y=S(t)},\;
t>0,
\tag{3.2}
\end{equation}

together with

\begin{equation}
\left[
\Phi^{-1}(\Phi^{-1})_x
-2x^3
-\frac{\lambda}{3}(t+a)^{-2}x^{-3}
\right]_{x\,|\,y=0}
=
H_0(t+a)^k,
\qquad
t>0.
\tag{3.3}
\end{equation}

Under the reciprocal transformation (2.4), the relation (2.5)$_1$ yelds $\Phi = 1/x_y$,   
so that with the symmetry reduction relation (2.7) wherein $m=-1/3, n=1/3$
there results

\begin{equation}
\Phi = (t+a)^{2/3} / \Psi'(y/(t+a)^{1/3}). 
\tag{3.4}\end{equation}
with
\begin{equation}
x = (t+a)^{-1/3} \Psi(y/(t+a)^{1/3}).
\tag{3.5}\end{equation}
Insertion of (3.4) into the moving boundary condition (3.2)$_1$ results in reduction to
\begin{equation}
(t+a)^{-1} \left[ \Psi''(\xi) - 2\Psi^3(\xi) - \frac{\lambda}{3}\Psi^{-3}(\xi) \right]_{\xi=\gamma} = L_m S^i \dot{S}
\tag{3.6}\end{equation}
so that $i=-1$ together with
\begin{equation}
L_m=3\left[\Psi''(\gamma)-2\Psi^3(\gamma)-\lambda/3\,\Psi^{-3}(\gamma)\right]
=\gamma\Psi(\gamma)
\tag{3.7}
\end{equation}
on application of the canonical Ermakov-Painlev\'e II reduction (2.9) corresponding
to $\zeta=0$. The boundary condition (3.2)$_2$ on use of the relation (3.5) requires
that $j=-1$ together with
\begin{equation}
P_m=\gamma\Psi(\gamma)
\tag{3.8}
\end{equation}
while the boundary requirement (3.3) yields
\begin{equation}
\Psi''(0)-2\Psi^3(0)-\lambda/3\,\Psi^{-3}(0)=H_0
\tag{3.9}
\end{equation}
together with $k=-1$. Accordingly, $H_0=0$ by virtue of the Ermakov-Painlev\'e II
reduction (2.9) with $\zeta=0$ and $\xi=0$.

In conclusion, it is remarked that Ermakov-Painlev\'e II symmetry
reduction has recently been applied in \cite{29} to solve a class of nonlinear 
moving boundary problems for a novel extension of the modified Kadomtsev-
Petviashvili equation. Moving boundary problems for a $2+1$-dimensional
extended Dym equation have been solved via Painlev\'e II symmetry
reductions in \cite{30}.

\subsection*{Acknowledgements}

S.C.\ acknowledges:
\begin{itemize}
\item Gr. Roma1, IV - Mathematical Methods in NonLinear Physics (MMNLP), National Institute for Nuclear Physics (I.N.F.N.), Rome, Italy;
\item Dipartimento di Scienze di Base e Applicate per l'Ingegneria,
Università di Roma  ``La Sapienza'', Rome, Italy;
\item Gruppo Nazionale di Fisica Matematica (G.N.F.M- I.N.D.A.M.), Italy;
\item Regione Lazio,   project CTE, Italy
\end{itemize}
 for supporting the present research activity.

\end{document}